\documentclass[manuscript,screen]{acmart}
\AtBeginDocument{%
  \providecommand\BibTeX{{%
    \normalfont B\kern-0.5em{\scshape i\kern-0.25em b}\kern-0.8em\TeX}}}

\copyrightyear{2022}
\acmYear{2022}
\setcopyright{acmcopyright}
\acmConference[RecSys '22]{Sixteenth ACM Conference
on Recommender Systems}{September 18--23, 2022}{Seattle, WA, USA}
\acmBooktitle{Sixteenth ACM Conference on Recommender Systems (RecSys '22),
September 18--23, 2022, Seattle, WA, USA}
\acmPrice{15.00}
\acmDOI{10.1145/3523227.3551481}
\acmISBN{978-1-4503-9278-5/22/09}

\usepackage{booktabs}
\usepackage{multirow}
\usepackage{xcolor}
\usepackage{subfig}
\usepackage{arydshln}

\let\savedegree\bigtimes
\let\bigtimes\relax
\usepackage{mathabx}
\let\bigtimes\savedegree

\usepackage{xspace}

\definecolor{ao(english)}{rgb}{0.0, 0.5, 0.0}

\newcommand{\ehsan}[1]{\textcolor{cyan}{{\bf [Ehsan: }{\em #1}{\bf ]}}}

\newcommand{\eg}{e.g., }
\newcommand{\ie}{i.e., }

\newcommand{\goso}{\texttt{GeoSoCa}\xspace}
\newcommand{\lore}{\texttt{LORE}\xspace}

\newcommand{\partitle}[1]{\vspace{2mm}\noindent\textbf{#1}}

\begin{document}


\title{Exploring the Impact of Temporal Bias in Point-of-Interest Recommendation}

\author{Hossein A.~Rahmani}
\affiliation{%
  \institution{University College London}
  \city{London}
  \country{United Kingdom}}
\email{h.rahmani@ucl.ac.uk}

\author{Mohammadmehdi Naghiaei}
\affiliation{%
  \institution{University of Southern California}
  \city{California}
  \country{USA}}
\email{naghiaei@usc.edu}

\author{Ali Tourani}
\affiliation{%
  \institution{University of Luxembourg}
  \city{Luxembourg City}
  \country{Luxembourg}}
\email{ali.tourani@uni.lu}

\author{Yashar Deldjoo}
\affiliation{%
  \institution{Polytechnic University of Bari}
  \city{Bari}
  \country{Italy}}
\email{deldjooy@acm.org}

\renewcommand{\shortauthors}{H.~A.~Rahmani, M.~Naghiaei, A.~Tourani, Y.~Deldjoo}

\begin{abstract}
Recommending appropriate travel destinations to consumers based on contextual information such as their check-in time and location is a primary objective of Point-of-Interest (POI) recommender systems. However, the issue of contextual bias (i.e., how much consumers prefer one situation over another) has received little attention from the research community. This paper examines the effect of temporal bias, defined as the difference between users' check-in hours, leisure vs.~work hours, on the consumer-side fairness of context-aware recommendation algorithms.
We believe that eliminating this type of temporal (and geographical) bias might contribute to a drop in traffic-related air pollution, noting that rush-hour traffic may be more congested. To surface effective POI  recommendation, we evaluated the sensitivity of state-of-the-art context-aware models to the temporal bias contained in users' check-in activities on two POI datasets, namely  Gowalla and Yelp. The findings show that the examined context-aware recommendation models prefer one group of users over another based on the time of check-in and that this preference persists even when users have the same amount of interactions.
\end{abstract}



\maketitle

\section{Introduction}
\label{sec:intro}
Recommender systems (RSs) are prevalent in the travel industry and the Point-of-Interest (POI) sector, helping users identify new locations and providing businesses with a means to attract new customers \cite{sanchez2022point,adomavicius2011context}. A POI recommender system may provide more tailored recommendations to users by exploiting \textbf{rich contextual information} such as the time and location of the users and POIs \cite{verbert2012context,raza2019progress,rahmani2022systematic}. The effectiveness of such systems is primarily assessed using conventional ranking-based metrics based on historical interactions.

Recently, there has been an increased awareness that such an approach fails to consider critical aspects of recommendation, such as fairness in one- and two-sided marketplaces~\cite{beutel2019fairness,burke2018balanced,melchiorre2020personality}. The primary technique for analyzing fairness in RSs is to provide equitable recommendation quality across groups of users or items, where an attribute define groups. In the literature, this concept is referred to as \textit{group fairness} \cite{deldjoo2022survey,naghiaei2022unfairness}. Its core tenet is that systems should not exhibit discriminatory performance against underprivileged groups as defined based on demographic attributes (\eg gender, race, age) or interaction-oriented qualities (\eg level of activity, mainstreamness, popularity) \cite{li2021user,kowald2020unfairness}. Although these attributes allow us to examine the system's fairness in related groups, we believe that they could be considered generic fairness, which does not fully explain the (un)fairness of the recommendation within a specific domain, such as the POI domain. Therefore, this research focuses on consumer-side fairness and underlines the need to analyze the group fairness of POI recommendations using in-domain and context-dependent knowledge.

To accomplish this, we describe two distinct user groups based on their interest in check-in times leveraging the temporal context and show how this difference favors one group over another. As a motivating example, we display the histograms of two users' check-ins (taken from the Gowalla dataset) and the recommendation quality offered to these individuals in Fig.~\ref{fig:user_checkins_hist}. Both users' profiles have a relatively equal number of interactions (891 vs.~938) and are from a similar demographic group. However, they vary significantly in their interest in check-in times. Alex has more check-ins during business hours, while John has more during leisure hours (see definitions in Section~\ref{sec:analysis}). The performance (evaluated by Precision, Recall, and nDCG) of the recommendations offered to these two individuals is the polar opposite. We claim that this temporal bias, which could be correlated with traffic congestion in metropolitan areas, might result in higher levels of air pollution from vehicle emissions and higher concentrations of pollutants. To remove or diminish such potential negative effects, the system designer and city policymakers may wish to assess and reduce the inequity in the geographical and temporal patterns of POI-recommendations, hereby referred to as \textbf{temporal unfairness}, which we believe can be associated with the measured levels of traffic-related air pollution.\footnote{Note the distinction between \textbf{temporal bias} and \textbf{temporal unfairness}. The former is measured solely by accounting user check-in activities while the latter is based on the suggestions provided by POI recommenders.} Toward this goal, we formulate the research questions according to the following dimensions:

\begin{figure*}
    \centering
    \includegraphics[scale=0.65]{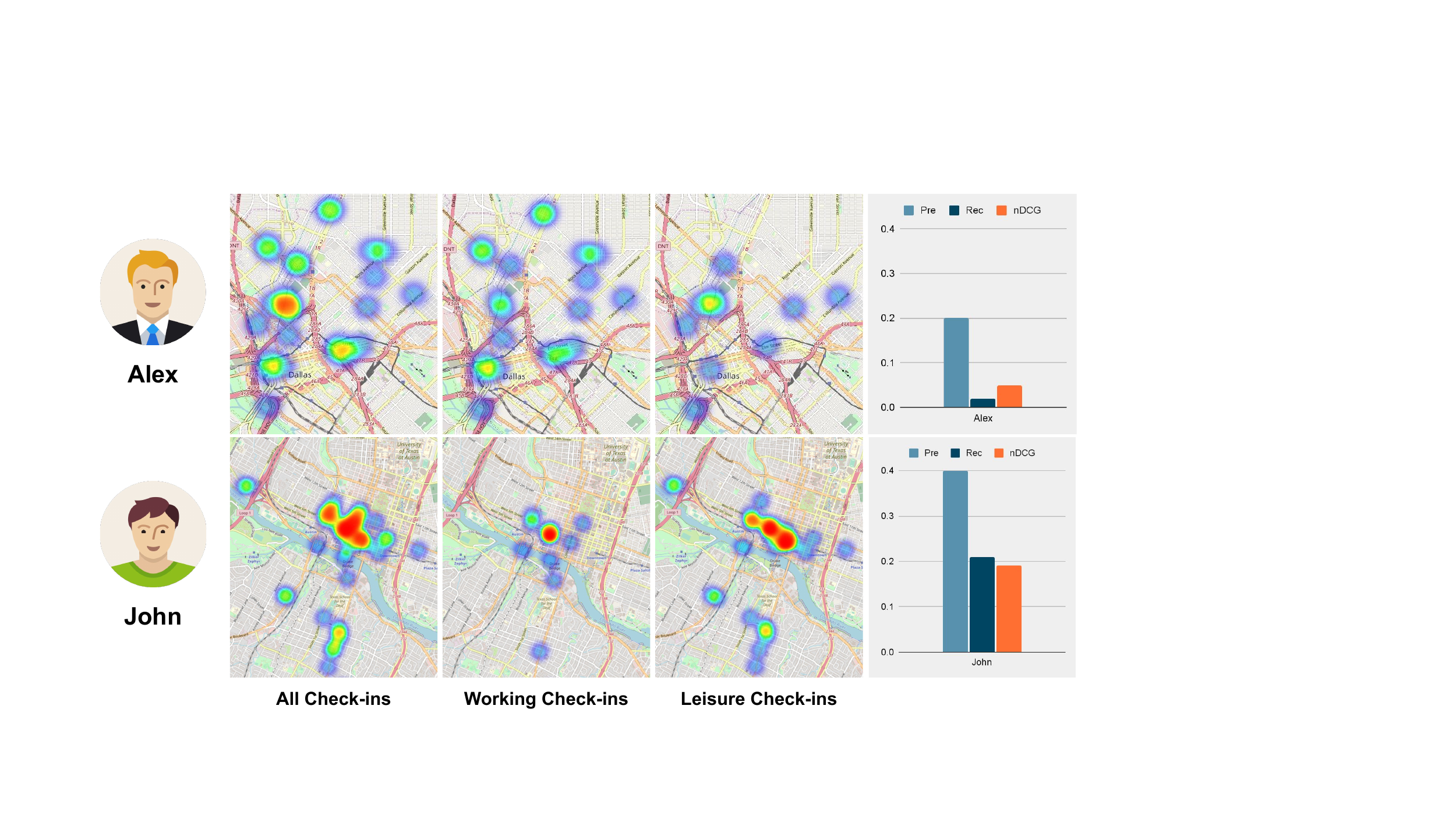}
    \caption{The spatio-temporal activity centers of two typical users in the Gowalla dataset. Alex and John both have the same number of check-ins, but their check-in patterns are distinct, and they received recommendations with varying degrees of quality.}
    \label{fig:user_checkins_hist}
\end{figure*}

\begin{itemize}
    \item \textbf{RQ1:} To what extent do the users or groups of users exhibit a distinct temporal check-in pattern?
    \item \textbf{RQ2:} How well do state-of-the-art (SoTA) context-aware POI models perform with respect to users displaying differing temporal behaviors?
    \item \textbf{RQ3:} Could context fusion methods (for instance, using sum rules instead of product rules~\cite{rahmani2022role}) allow for a more accurate representation of temporal behavior? Would modifying the fusion method or contextual weights improve fairness?
\end{itemize}

In the following, we investigate \textbf{RQ1} in Section \ref{sec:analysis}. Then, we discuss our experimental setup and evaluation method in Section \ref{sec:results}. We examine \textbf{RQ2} and \textbf{RQ3} and discuss the findings of our study in Section~\ref{sec:results} and conclude the paper in Sections \ref{sec:conclusion}.
Finally, to enable reproducibility of the experiment, we have made our codes open source.\footnote{\url{https://github.com/rahmanidashti/ContextsFair}}
\section{Related Work and Background}
\label{sec:relatedwork}
A context-aware RS should provide a user with recommendations taking into consideration the user’s current context. Due to the importance of contextual information in POI recommendation, several researchers have tried to investigate and explore the impact of contextual information for POI recommendation task.
For instance, \citet{rahmani2020joint} analyze the pattern of users' check-in based on geographical and temporal information in POI recommendation and show that users' behavioral check-ins vary across temporal states.
\citet{Aliannejadi2018} take a different approach to user behavior mining by fusing diverse contextual information, such as location and user comments, in order to overcome insufficient preference information and thereby improve POI recommendation accuracy. \citet{MiglioriniQCB19} investigate the role of context in recommending a sequence of activities to groups of users. In their approach, the authors demonstrated that by introducing a multi-objective optimization method, the contextual information has an impact on user preferences and results in enhanced recommendations. In \cite{yu2021leveraging}, a context- and preference-aware approach is developed for recommending POIs based on the contextual influences and preferences of the users. Accordingly, the relationship between users and points of interest (POIs) is revealed in light of the anticipated score of locales. In contrast, we focus our attention on the consumer-side fairness and emphasize on the relevance of employing \textbf{in-domain} fairness scenarios when examining group fairness in the POI recommendation domain. In particular, we aim to look at the bias posed from the temporal context perspective.
\section{Analyzing the Context-aware Behavior (RQ1)}
\label{sec:analysis}
In this section, we demonstrate the temporal behavior of users through data-driven observational analysis.

\partitle{Datasets.}
In this paper, we utilize two real-world well-known check-in datasets provided by \citet{liu2017experimental}, namely, \textit{Gowalla} and \textit{Yelp}. The Gowalla dataset is collected from February 2009 to October 2010. Following \cite{liu2017experimental} and \cite{rahmani2019lglmf}, we preprocessed the Gowalla dataset by removing cold users, \ie users who visited locations for less than $15$ check-ins. We also excluded POIs of less than $10$, which may cause a spam error on the model. The Yelp dataset is provided by the Yelp dataset challenge\footnote{\url{https://www.yelp.com/dataset}} round 7 (access date: Feb 2016) in 10 metropolitan areas across two countries. We also preprocessed the Yelp dataset and removed users who have less than $10$ visited locations and POIs with less than $10$ visits. Table \ref{tbl:datasets} shows the statistics of the datasets.

\begin{table*}
  \caption{Characteristics of the datasets: $\left| \mathcal{U} \right|$ is the number of users, $\left| \mathcal{P} \right|$ is the number of POIs, $\left| \mathcal{C} \right|$ is the number of check-ins, $\left| \mathcal{C}_u \right|$ is the number of unique check-ins, $\left| \mathcal{S} \right|$ is the number of social link, $\left| \mathcal{G} \right|$ is the number of category, $\frac{\left| \mathcal{C} \right|}{\left| \mathcal{U\times{P}} \right|}$ is the density.}
  \centering
  \label{tbl:datasets}
  \begin{tabular}{cccccccccc}
    \toprule
    \textbf{Dataset} & $\left| \mathcal{U} \right|$ & $\left| \mathcal{P} \right|$ & $\left| \mathcal{C} \right|$ & $\left| \mathcal{C}_u \right|$ & $\left| \mathcal{S} \right|$ & $\left| \mathcal{G} \right|$ & $\frac{\left| \mathcal{C} \right|}{\left| \mathcal{U} \right|}$ & $\frac{\left| \mathcal{C} \right|}{\left| \mathcal{P} \right|}$ & $\frac{\left| \mathcal{C} \right|}{\left| \mathcal{U\times{P}} \right|}$ \\
    \midrule
    \textbf{Yelp} 
    & 7,135 & 16,621 & 1,137,521 & 285,608 & 46,778 & 595 & 159.42 & 68.43 & 0.0095 \\
    \midrule
    \textbf{Gowalla} 
    & 5,628 & 31,803 & 620,683 & 378,968 & 46,001 & - & 110.28 & 19.51 & 0.0034 \\
    \bottomrule 
  \end{tabular}
\end{table*}

\begin{figure*}[!tbp]
  \centering
  \subfloat[Histogram of timestamp of check-ins (\textcolor{red}{Red} shows working hours while \textcolor{ao(english)}{green} indicates leisure hours)]
  {\includegraphics[scale=0.235]{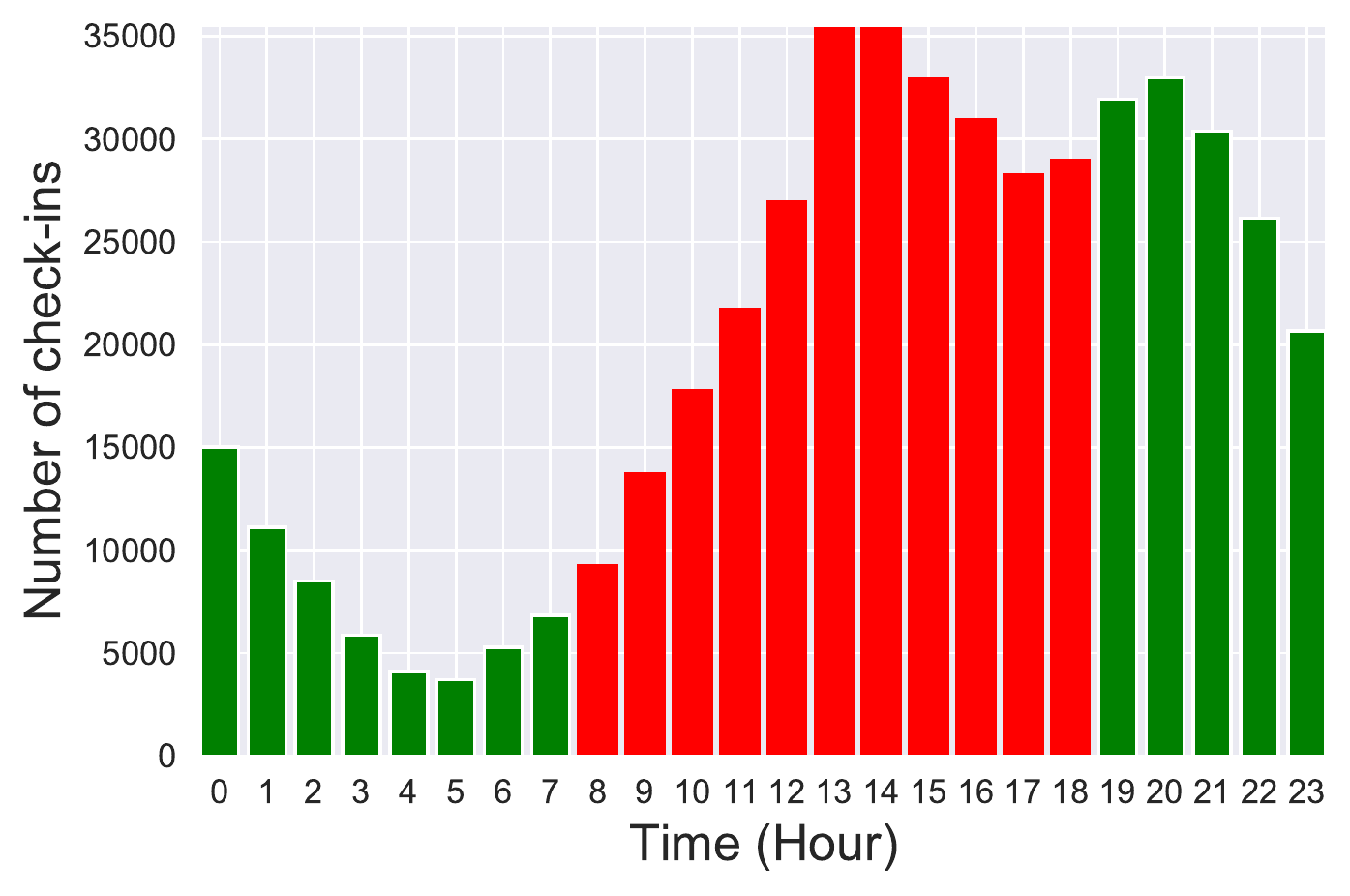}\label{fig:checkins_time_hist}}
  \qquad
  \subfloat[No.~of check-ins in leisure time vs.~no. of check-ins working time with x=y line]
  {\includegraphics[scale=0.235]{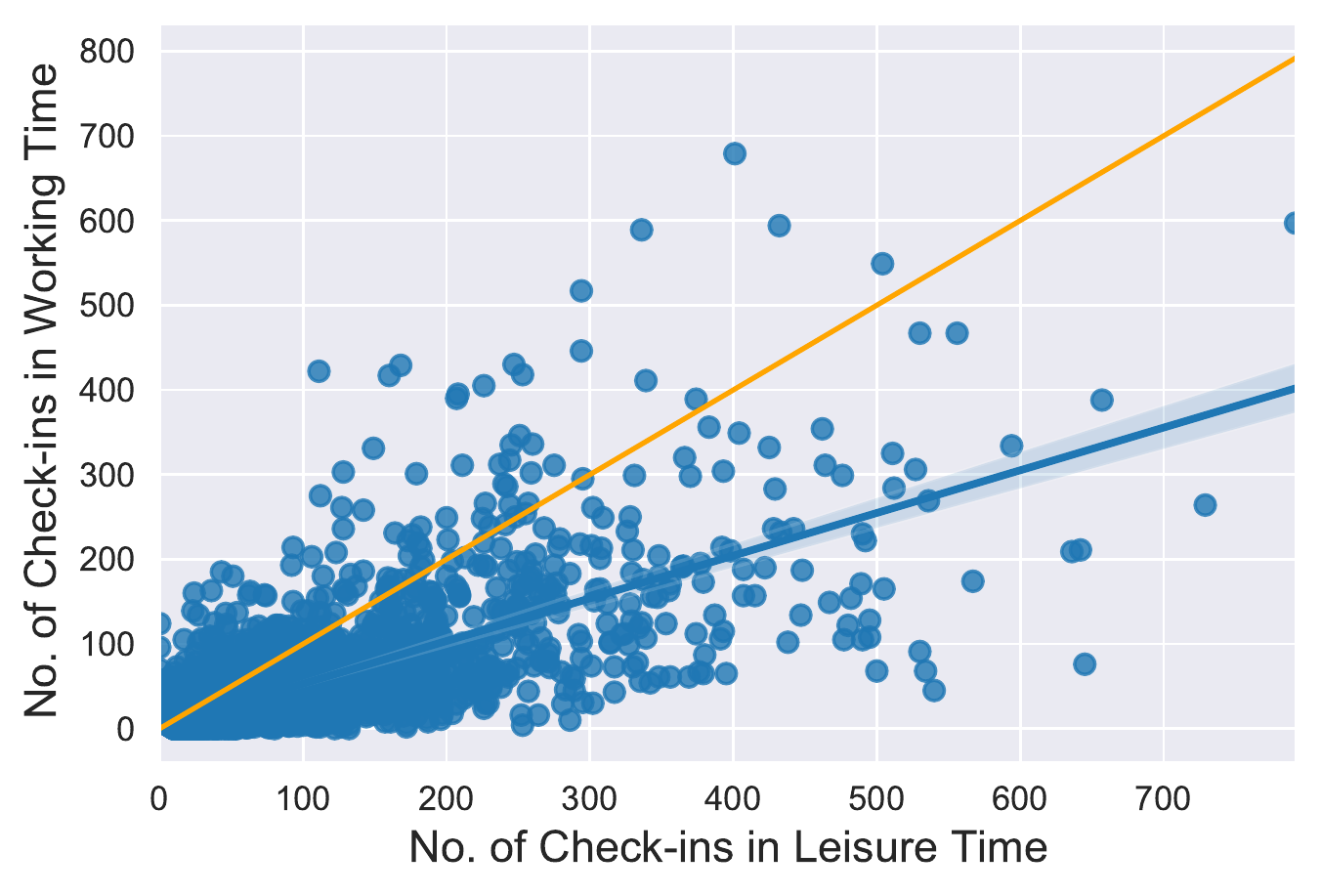}\label{fig:leisure_working}}
  \qquad
  \subfloat[Percentage of leisure time check-ins vs. user profile size]{\includegraphics[scale=0.235]{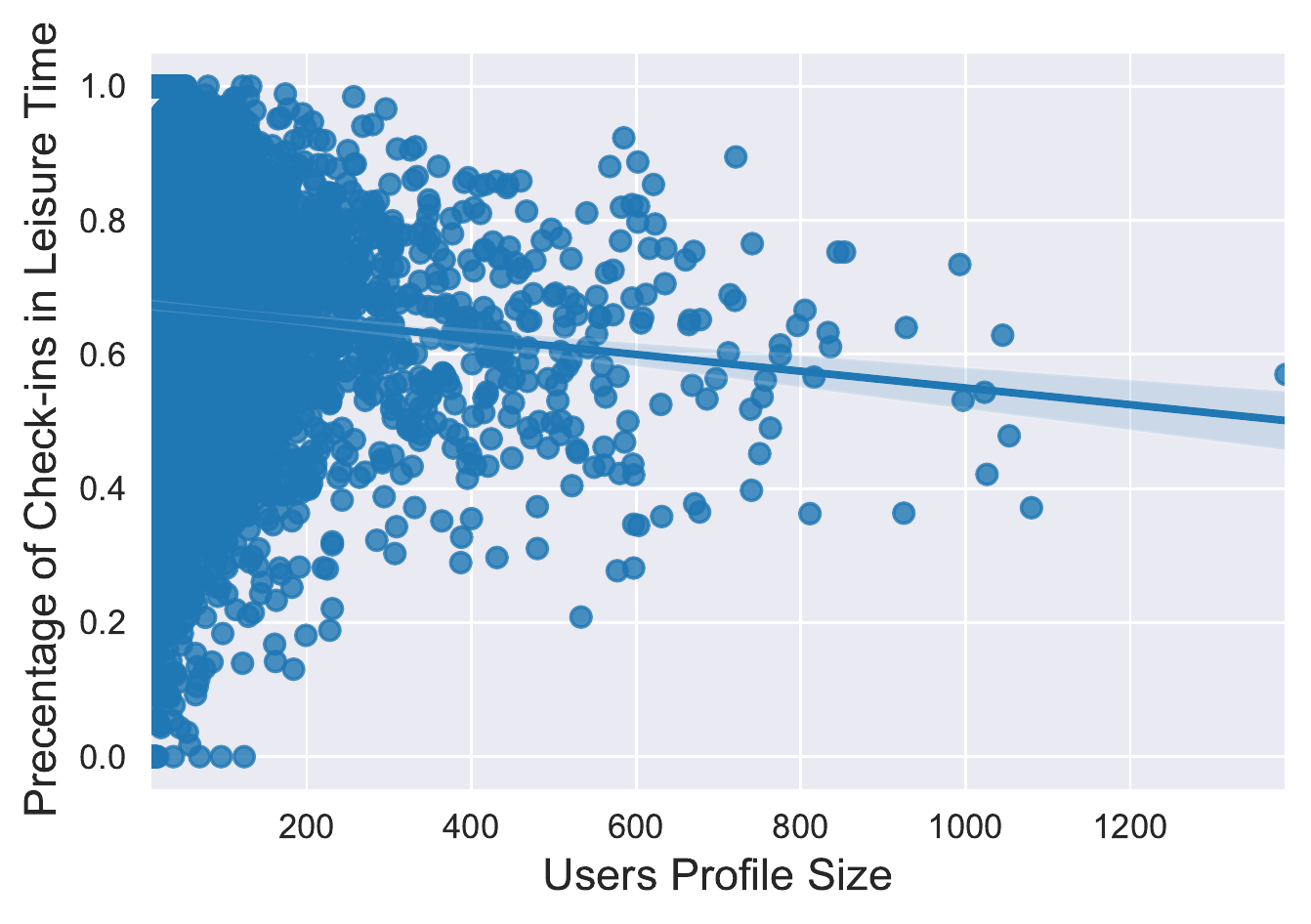}
  \label{fig:ratio_leisure_profile}}
  \qquad
  \subfloat[Percentage of working time check-ins vs. user profile size ]{\includegraphics[scale=0.235]{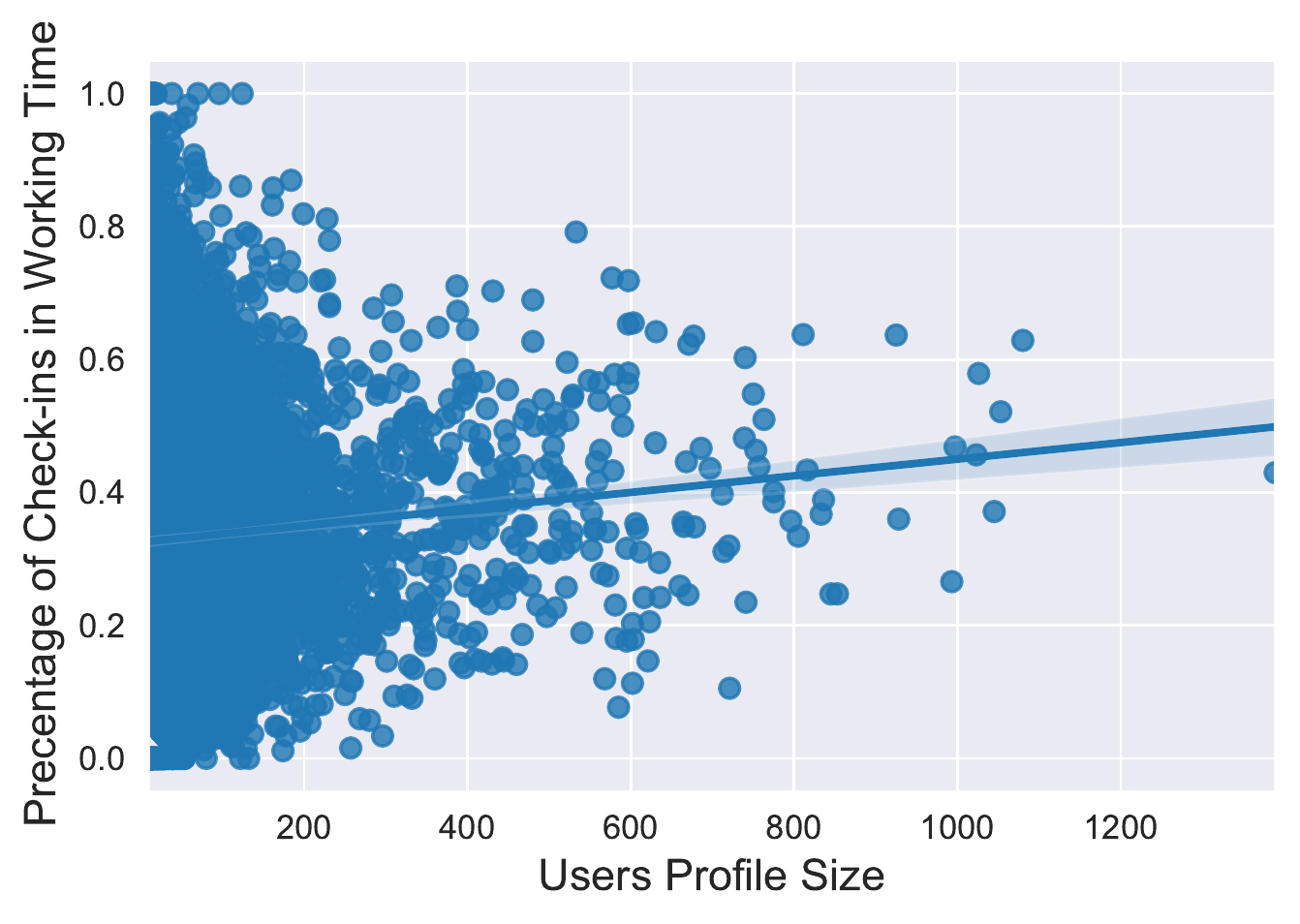}
  \label{fig:ratio_working_profile}}
  \caption{Gowalla dataset check-ins distribution and users' temporal check-ins behavior.}
  \label{fig:data_dis}
\end{figure*}

\partitle{Users Check-in Distribution.} Fig.~\ref{fig:checkins_time_hist} shows histogram of check-in  timestamps. As can be seen, it is evident that the distribution of check-ins are not homogeneous and skews toward the afternoons and evenings hours. For a more detailed analysis of the temporal behavior, following \citet{rahmani2020joint} and \citet{liu2013personalized}, we further categorize the check-ins based on their timestamps into two categories: working time and leisure time. Check-ins between 8 AM and 6 PM categorize as working time check-ins, whereas the remainder falls under the second category.\footnote{Although, the idea should be trivially extendable to other selections exhibiting non-homogeneous pattern}
Fig.~\ref{fig:leisure_working} depicts the positive correlation between the number of check-ins during leisure and working hours for each user. This is expected due to the fact that every interaction may occur in the course of working or leisure time, and hence, the likelihood of seeing more of both types of interactions increases as the user's profile size increases. Note that the slope of this correlation is less than one (the slope of $X=Y$ is illustrated in \textcolor{orange}{orange} in Fig.~\ref{fig:leisure_working}), suggesting that active users have non-uniform density of check-in during working and leisure hours. Hence, our next objective is to analyze the users' temporal behavior in relation to check-ins. Particularly, we are interested in determining whether certain users prefer leisure hours check-in over working hours and vice versa.

\partitle{Temporal Bias in Users' Behaviour.}
In order to provide further clarification of users' tastes, Figs.~\ref{fig:ratio_leisure_profile} and~\ref{fig:ratio_working_profile} illustrate the correlation between the ratio of leisure and working check-ins within the user profile and the size of their profile, respectively. Interestingly, the number of leisure time check-ins is negatively correlated with the size of the user profile, whereas the number of working hours check-ins is positively correlated. This implies that users with large profiles are more likely to check-in during working hours.  
Based on our analysis results, it is apparent that users have a different tendency toward check-ins at different times of the day (i.e., during working hours or leisure hours). To capture and analyze this behavior of users and examine the effects of recommendation quality on them separately, we divide users into two groups as follows: 

\begin{itemize}
     \item \textbf{Leisure-focused users}: After sorting users according to the percentage of leisure time check-ins in their profiles, we refer to the top $20\%$ of this list as leisure-focused users.
     \item \textbf{Working-focused users}: We consider the $20\%$ users at the bottom of the sorted list as working-focused, i.e., users with the lowest number of leisure time check-ins
\end{itemize}

Table~\ref{tbl:user_groups} shows the characteristic and interaction behavior of both groups. As one may notice, the characteristics of the two groups are quite similar, with leisure-focused users displaying slightly higher levels of interaction and popularity consumption.
In what follows, we select two SoTA POI context-aware algorithms and investigate their performance with respect to the two user groups. We observed similar patterns in the Yelp dataset, though the analysis is not included due to space consideration. 

\begin{table*}
\caption{Data characteristics of the user groups.}
  \centering
  \label{tbl:user_groups}
\begin{tabular}{llcccc}
\toprule
Dataset & Group & No.~of Check-ins & Avg.~Popularity Consumption & Avg.~Activity Level & No.~of users \\ \midrule
Gowalla & Working-focused & 246373 & 0.535 & 218.80 & 1125 \\
Gowalla & Leisure-focused & 273610 & 0.552 & 242.99 & 1125 \\
\bottomrule
\end{tabular}
\end{table*}

\section{Results and Discussion}
\label{sec:results}

\begin{table*}
\centering
\caption{Recommendation performance of GeoSoCa and LORE models with different fusion methods on Gowalla and Yelp datasets for all, leisure-focused, and working-focused users. The evaluation metrics are calculated based on the top-10 and top-20 predictions in the test set. Our best results are highlighted in \textbf{bold}. $\%\Delta$ values denote the percentage of relative improvement compared to $\bigodot$ fusion and Acc./Unf. shows the ratio of overall nDCG to $\Delta$nDCG.}
\label{tbl:results}
\resizebox{\columnwidth}{!}{%
\begin{tabular}{lclllllllcllllllllc}
\toprule
\multirow{2}{*}{\textbf{Model}} & \multirow{2}{*}{\textbf{Fusion}} & \multicolumn{8}{c}{\textbf{@10}} && \multicolumn{8}{c}{\textbf{@20}} \\
\cmidrule{3-10}\cmidrule{12-19}
 & & Pre & Rec & nDCG & nDCG$_L$ & nDCG$_W$ & $\Delta$nDCG $\downarrow$ & \%$\Delta$ & Acc./Unf. && Pre & Rec & nDCG & nDCG$_L$ & nDCG$_W$ & $\Delta$nDCG $\downarrow$ & \%$\Delta$ & Acc./Unf. \\
\midrule
\multicolumn{19}{c}{\textbf{Gowalla}}\\ \hline
\multirow{3}{*}{GeoSoCa} & $\bigodot$    & 0.0344 & 0.0329 & 0.0368 & 0.0679 & 0.0226 & 0.0453 & 0      & 0.8123 && 0.0274 & 0.0507 & 0.0313 & 0.0590 & 0.0187 & 0.0409 & 0      & 0.7652 \\
                         & $\bigoplus$   & 0.0335 & 0.0339 & 0.0354 & 0.061 & 0.02240 & 0.0386 & 0.1479 & 0.9170 && 0.0282 & 0.0557 & 0.0312 & 0.0560 & 0.0193 & 0.0367 & 0.1026 & 0.8501 \\
                         & $\bigboxplus$ & 0.0322 & 0.0326 & 0.0339 & 0.0579 & 0.0220 & 0.0359 & 0.2075 & \textbf{0.9442} && 0.0272 & 0.0538 & 0.0299 & 0.0530 & 0.0188 & 0.0342 & 0.1638 & \textbf{0.8742} \\ \hdashline
\multirow{3}{*}{LORE}    & $\bigodot$    & 0.0450 & 0.0404 & 0.0498 & 0.0973 & 0.0287 & 0.0686 & 0      & 0.7259 && 0.0349 & 0.0590 & 0.0413 & 0.0855 & 0.0226 & 0.0629 & 0      & 0.6565 \\
                         & $\bigoplus$   & 0.0345 & 0.0346 & 0.0363 & 0.0624 & 0.0226 & 0.0398 & 0.4198 & 0.9120 && 0.0292 & 0.0570 & 0.0321 & 0.0570 & 0.0195 & 0.0375 & 0.4038 & 0.856 \\
                         & $\bigboxplus$ & 0.0352 & 0.0355 & 0.0373 & 0.0632 & 0.0238 & 0.0394 & 0.4256 & \textbf{0.9467} && 0.0297 & 0.0579 & 0.0329 & 0.0575 & 0.0203 & 0.0372 & 0.4085 & \textbf{0.8844} \\ \hline
\multicolumn{19}{c}{\textbf{Yelp}}\\ \hline
\multirow{3}{*}{GeoSoCa} & $\bigodot$    & 0.0184 & 0.0215 & 0.0196 & 0.0519 & 0.0056 & 0.0463 & 0      & 0.4233 && 0.0151 & 0.0341 & 0.0169 & 0.0458 & 0.0048 & 0.0410 & 0      & 0.4121 \\
                         & $\bigoplus$   & 0.0165 & 0.0216 & 0.0173 & 0.0422 & 0.0070 & 0.0352 & 0.2397 & 0.4914 && 0.0150 & 0.0400 & 0.0160 & 0.0390 & 0.0070 & 0.0320 & 0.2195 & 0.5 \\
                         & $\bigboxplus$ & 0.0150 & 0.0204 & 0.0157 & 0.0367 & 0.0071 & 0.0296 & 0.3606 & \textbf{0.5304} && 0.0140 & 0.0385 & 0.0148 & 0.0343 & 0.0070 & 0.0273 & 0.3341 & \textbf{0.5421} \\ \hdashline
\multirow{3}{*}{LORE} & $\bigodot$       & 0.0183 & 0.0203 & 0.0196 & 0.0505 & 0.0051 & 0.0454 & 0      & 0.4317 && 0.0157 & 0.0335 & 0.0174 & 0.0468 & 0.0043 & 0.0425 & 0      & 0.4094 \\
                         & $\bigoplus$   & 0.0162 & 0.0215 & 0.0169 & 0.0395 & 0.0069 & 0.0326 & 0.2819 & \textbf{0.5184} && 0.0145 & 0.0388 & 0.0155 & 0.0360 & 0.0066 & 0.0294 & 0.3082 & 0.5272 \\
                         & $\bigboxplus$ & 0.0155 & 0.0205 & 0.0159 & 0.0377 & 0.0066 & 0.0311 & 0.3149 & 0.5112 && 0.0139 & 0.0377 & 0.0147 & 0.0339 & 0.0064 & 0.0275 & 0.3529 & \textbf{0.5345}\\
\bottomrule
\end{tabular}
}
\end{table*}

In this section, we evaluate \textbf{RQ2} and \textbf{RQ3} by analyzing the performance of context-aware SoTA POI algorithms on the user groups defined earlier. 

\subsection{Evaluation Protocols and Models}
First, we discuss SoTA POI recommendation models used in our study as well as our evaluation protocols. We used two well-known and most cited context-aware POI recommendation models, \goso ~\cite{zhang2015geosoca} and \lore~\cite{zhang2014lore}.

\begin{itemize}
    \item \textbf{\goso:} The \goso~model takes into account three different contextual information, namely geographical, social, and categorical. \goso uses the three contextual models to develop a unified preference score $\goso(u,p)$ for the user $u$ on an unvisited POI $p$ according to $F_{{G}_{up}} \bigodot F_{{S}_{up}} \bigodot F_{{C}_{up}}$ where $F_{{G}_{up}}$, $F_{{S}_{up}}$, and $F_{{C}_{up}}$ show the user $u$'s preference on POI $p$ based on the geographical, social, and categorical component of \goso, respectively, and $\bigodot$ denotes the product operation.
    \item \textbf{\lore:} \lore takes advantage of the sequential influences on POI recommendation, which include geographic, temporal, and social contexts. \lore~fuses these context information by integrating their derived sequential probability with geographical probability and social rating of a user visiting a new location into a single score, $F_{{K}_{up}} \bigodot F_{{U}_{up}} \bigodot F_{{T}_{up}}$, where $F_{{K}_{up}}$ is the score of geographical context, $F_{{U}_{up}}$ is score of the social component, $F_{{T}_{up}}$ represents the temporal context, and $\bigodot$ is the product rule.
\end{itemize}

Further, three ranking-based evaluation metrics including Pre@$N$ (Precision at $N$), Rec@$N$ (Recall at $N$), and nDCG@$N$ with $N \in \{10, 20\}$ are used to evaluate the performance of the recommendation methods. We also use $\Delta$nDCG to evaluate the fairness according to two proposed user groups, \ie leisure-focused and working-focused. We partition each dataset into training data, validation data, and test data. For each user, we use the earliest 70\% check-ins as the training data, the most recent 20\% check-ins as the test data and the remaining 10\% as the validation data. 

\subsection{Temporal Unfairness of Context-aware POI Recommendation Models (RQ2)}
\begin{figure*}
  \centering
  \subfloat[Yelp]
  {
    \includegraphics[scale=0.3]{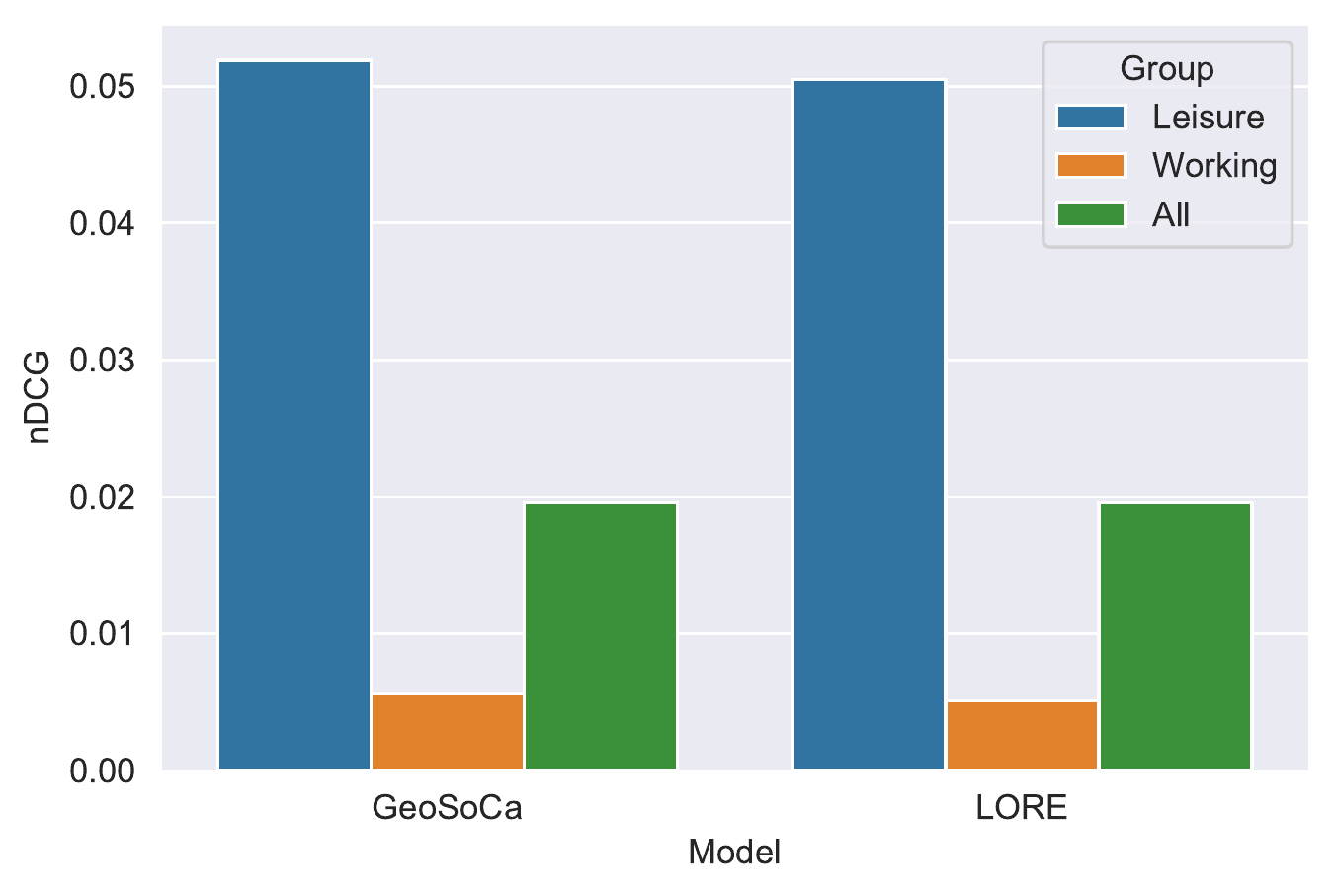}
    \label{fig:yelp_10}
  }
  \subfloat[Gowalla]
  {
    \includegraphics[scale=0.3]{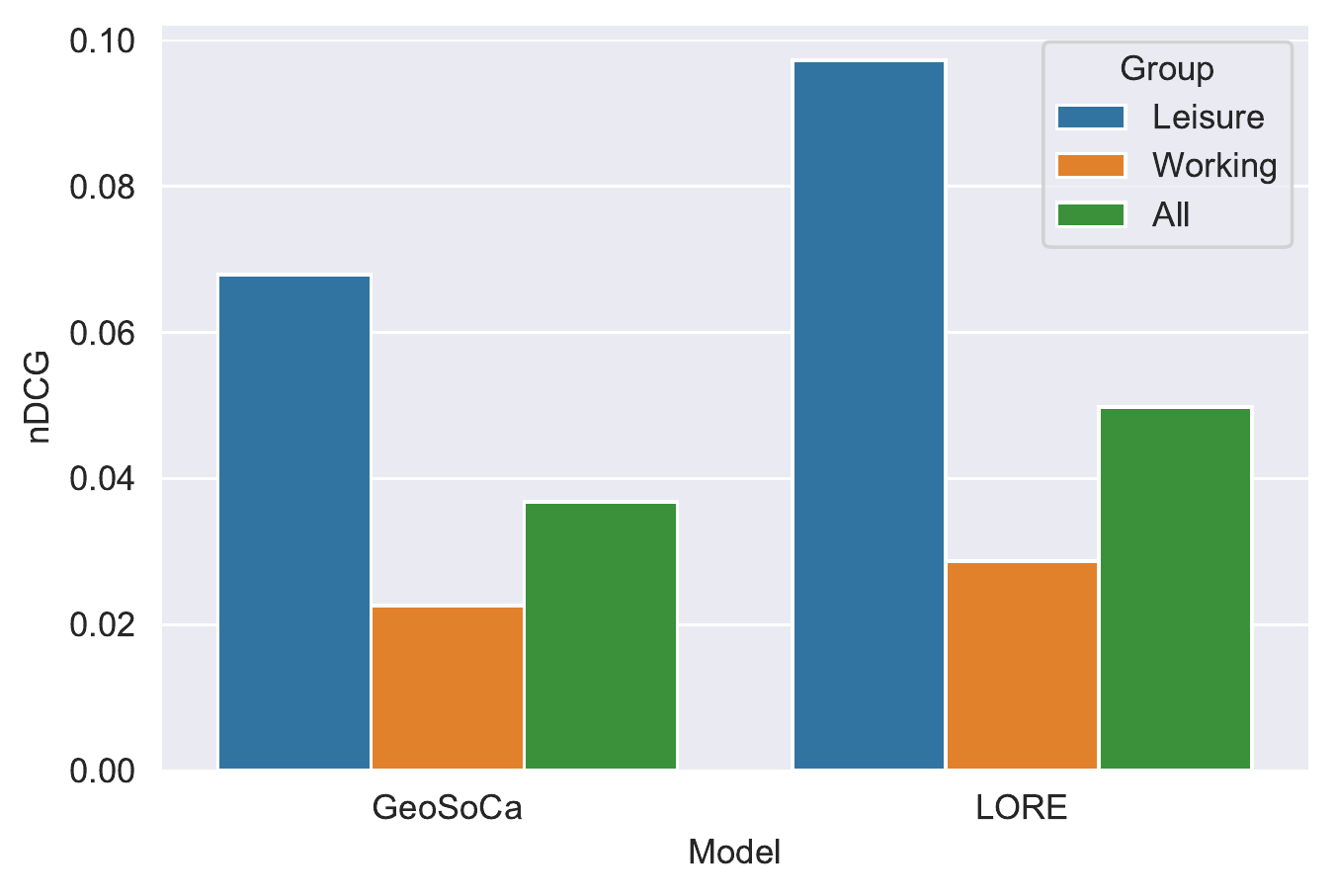}
    \label{fig:gowalla_10}
  }
    \subfloat[Fusion weights effects on fairness]
  {
    \includegraphics[scale=0.3]{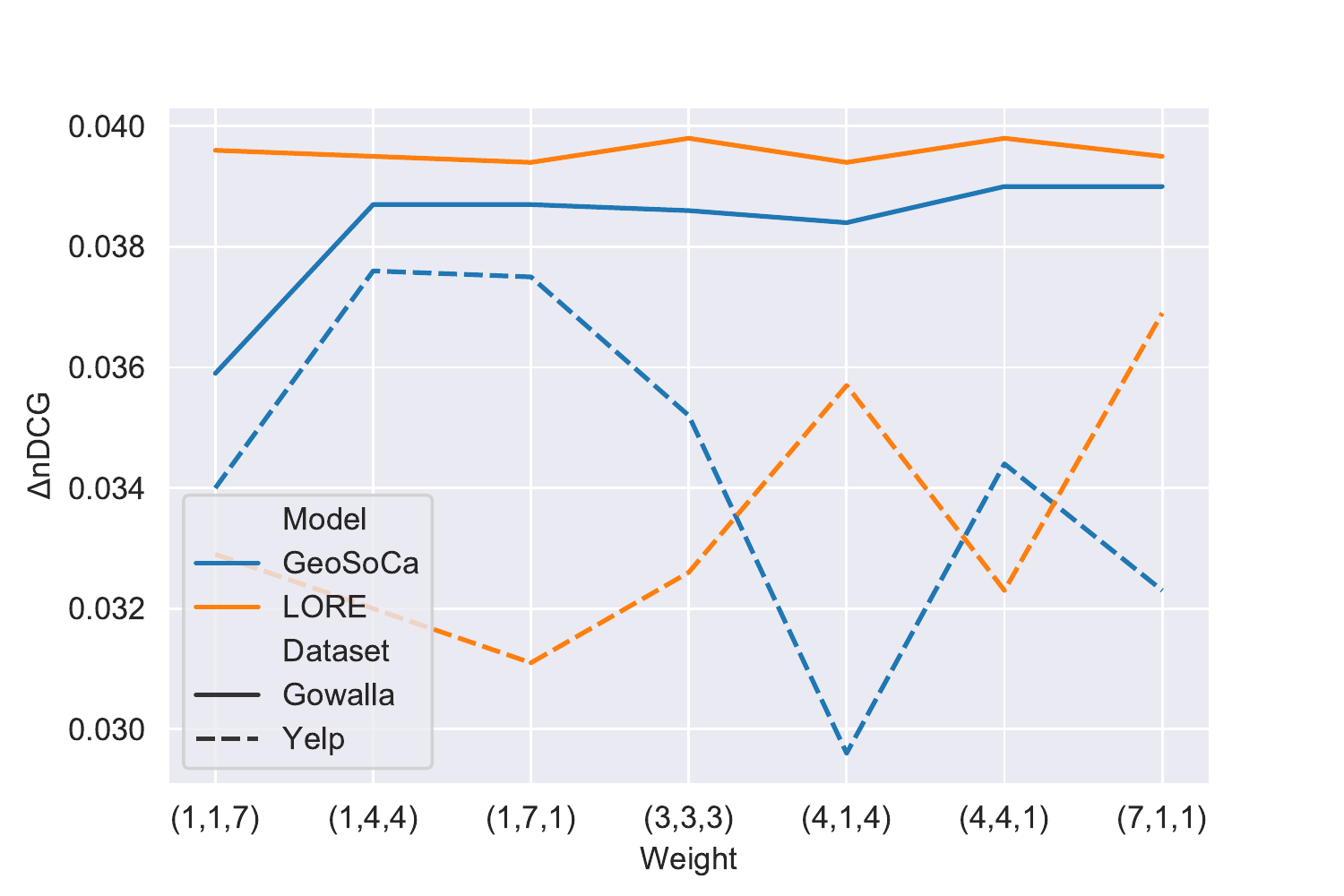}
    \label{fig:linechart}
  }
  \caption{Analysis of the unfairness of context-aware recommendations. Figs.~\ref{fig:yelp_10} and \ref{fig:gowalla_10} show the nDCG@10 values among different user groups and Fig.~\ref{fig:linechart} shows fusion weights effect on difference between nDCG@10 of leisure and working user groups.}
  \label{fig:barchart}
\end{figure*}

Figs.~\ref{fig:yelp_10} and \ref{fig:gowalla_10} display the nDCG values for recommendation algorithms among user groups and all users separately for the Yelp and Gowalla datasets. There can be seen that leisure-focused users benefit from a higher quality of recommendation for both models and datasets, even though the overall characteristics of leisure-focused users are similar to those of working-focused users, as shown in Table~\ref{tbl:user_groups}. For instance, the results in Table~\ref{tbl:results} on the Gowalla dataset show that nDCG values for the two models of (GeoSoCa, Lore) are (0.0368, 0.0498), (0.0679, 0.0973), and (0.0226, 0.0287) on leisure-focused users, working-focused users, and all respectively. This findings indicate that context-aware POI recommendations are capable of achieving high overall accuracy when incorporating contextual information, but are unable to capture the difference in check-in preferences of users, resulting in a disparity in quality between those belonging to different groups.

\subsection{Context Fusion Effect on Fairness (RQ3)}
\label{sec:fusioneffect}
Here we will investigate whether the fusion method for incorporating contextual information could play a role in mitigating temporal unfairness. In particular, we investigate the use of sum ($\bigoplus$) and weighted sum ($\bigboxplus$) instead of the product ($\bigodot$) rule initially used in \lore~and \goso~models to fuse contextual factors in a personalized fashion. A polynomial regression fusion model can be used to model the general fusion of existing POI recommendations:

\begin{definition}[Polynomial Regression Context Fusion]
Let $c_i$ refer to context $i$ (\eg geographical, social, etc.). The contextual information could be infused using a polynomial regression model according to:
\begin{small}
\begin{equation}
POIRec_{u,p} = \lambda_1 c_1 + \lambda_2 c_2 + \lambda_3 c_3 + \lambda_{12} c_1c_2 + \lambda_{13} c_1c_3 + \lambda_{23} c_2c_3 + \lambda_{123}c_1c_2c_3
\label{eq:ployregcontext}
\end{equation}
\end{small}
\noindent
in which $\lambda_j$ indicates the importance weight for the context $c_i$ learned by the model. Note that the product rule ($\bigodot$) would have $\lambda_j =0$ for all $j$ and $\lambda_{123}=1$. In the case of the sum ($\bigoplus$), $\lambda_1$, $\lambda_2$, and $\lambda_3$ are $1$ and the rest are $0$, while in the weighted sum ($\bigboxplus$), optimal values are assigned to them to maximize fairness criteria.
\qed
\label{def:polycontextfusion}
\end{definition}

\begin{figure*}
  \centering
  \subfloat[Gowalla | GeoSoCa]
  {
    \includegraphics[scale=0.2]{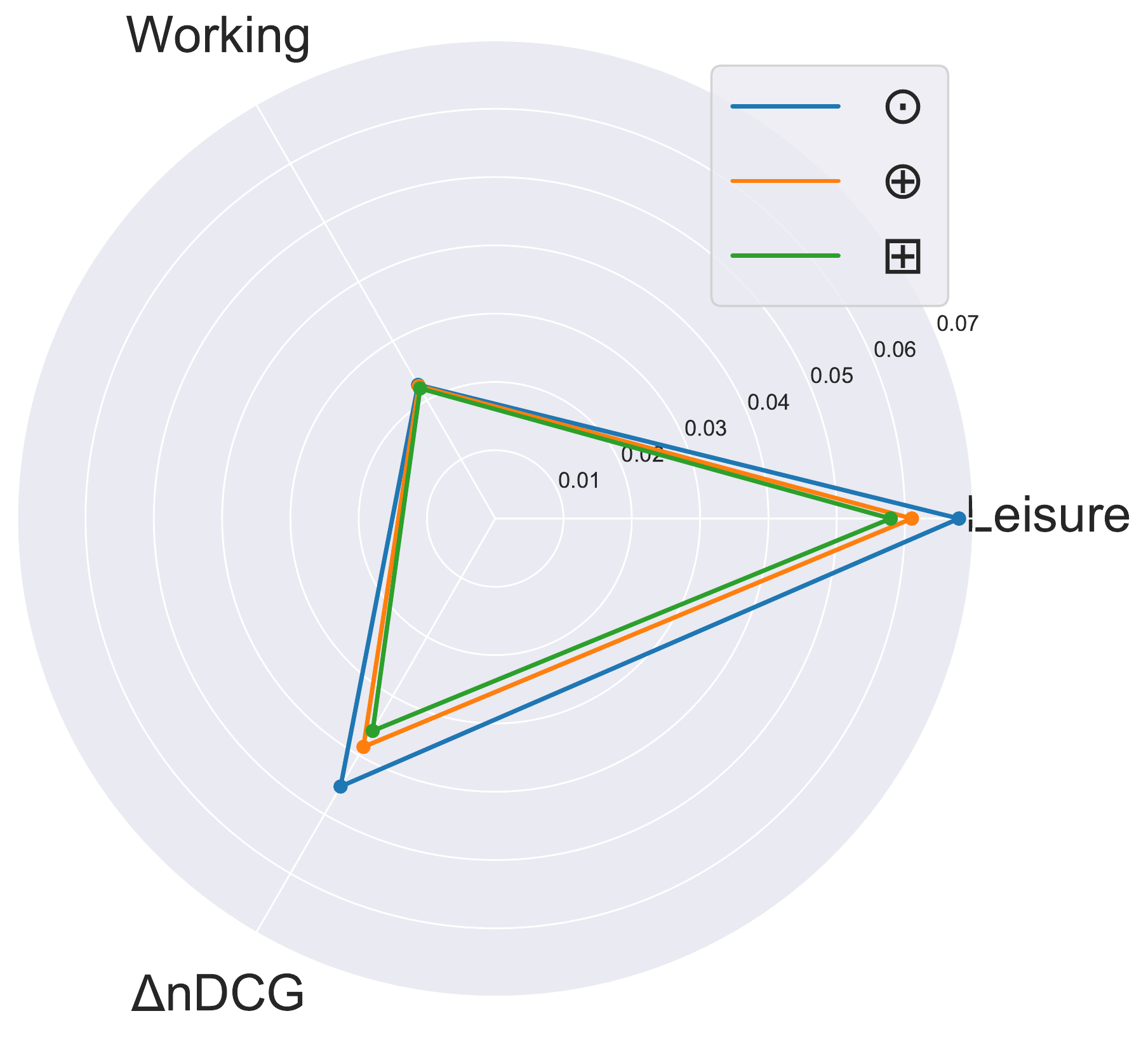}
    \label{fig:Gowalla_GeoSoCa_10}
  }
  \subfloat[Gowalla | LORE]
  {
    \includegraphics[scale=0.2]{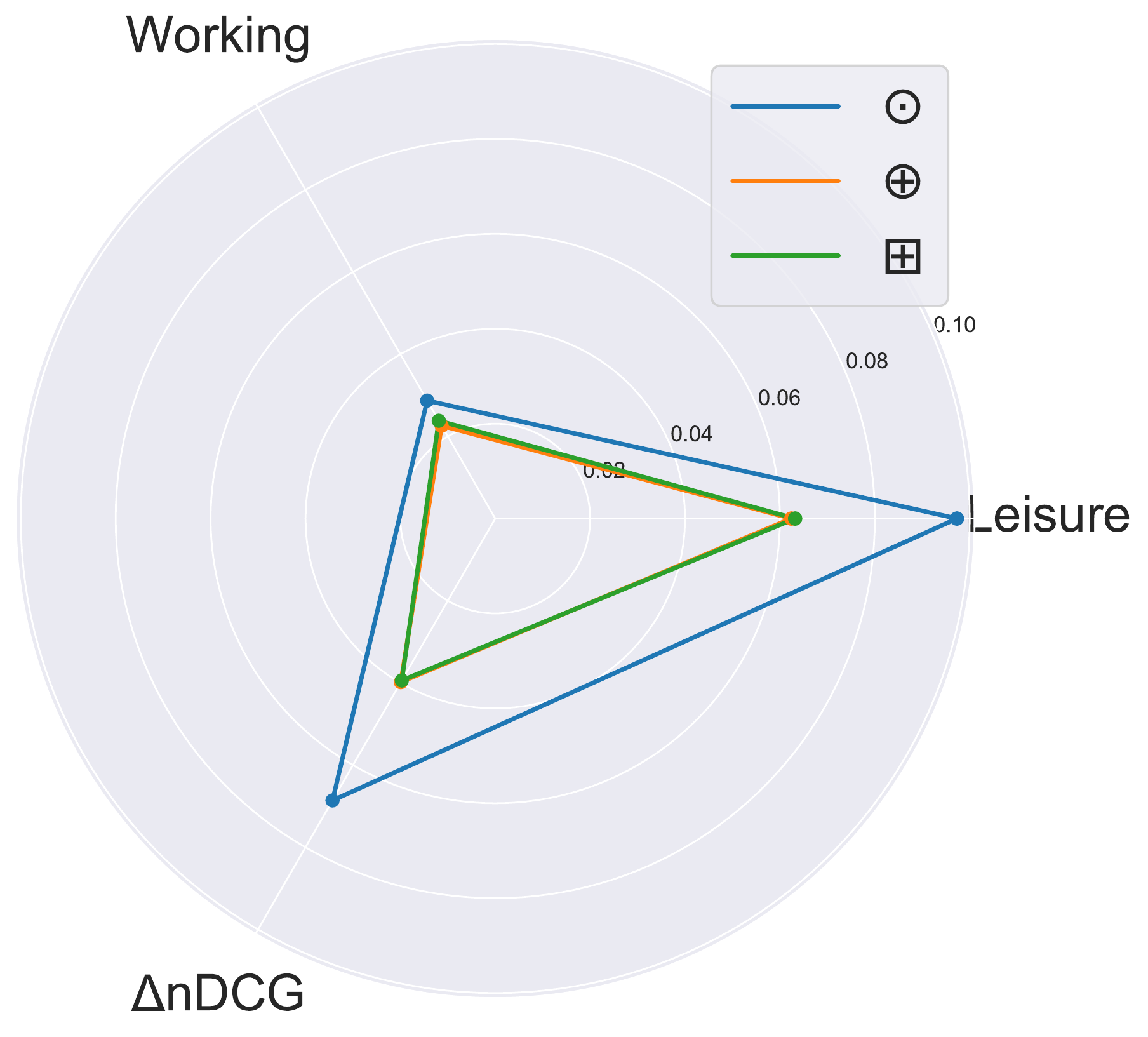}
    \label{fig:Gowalla_LORE_10}
  }
    \subfloat[Yelp | GeoSoCa]
  {
    \includegraphics[scale=0.2]{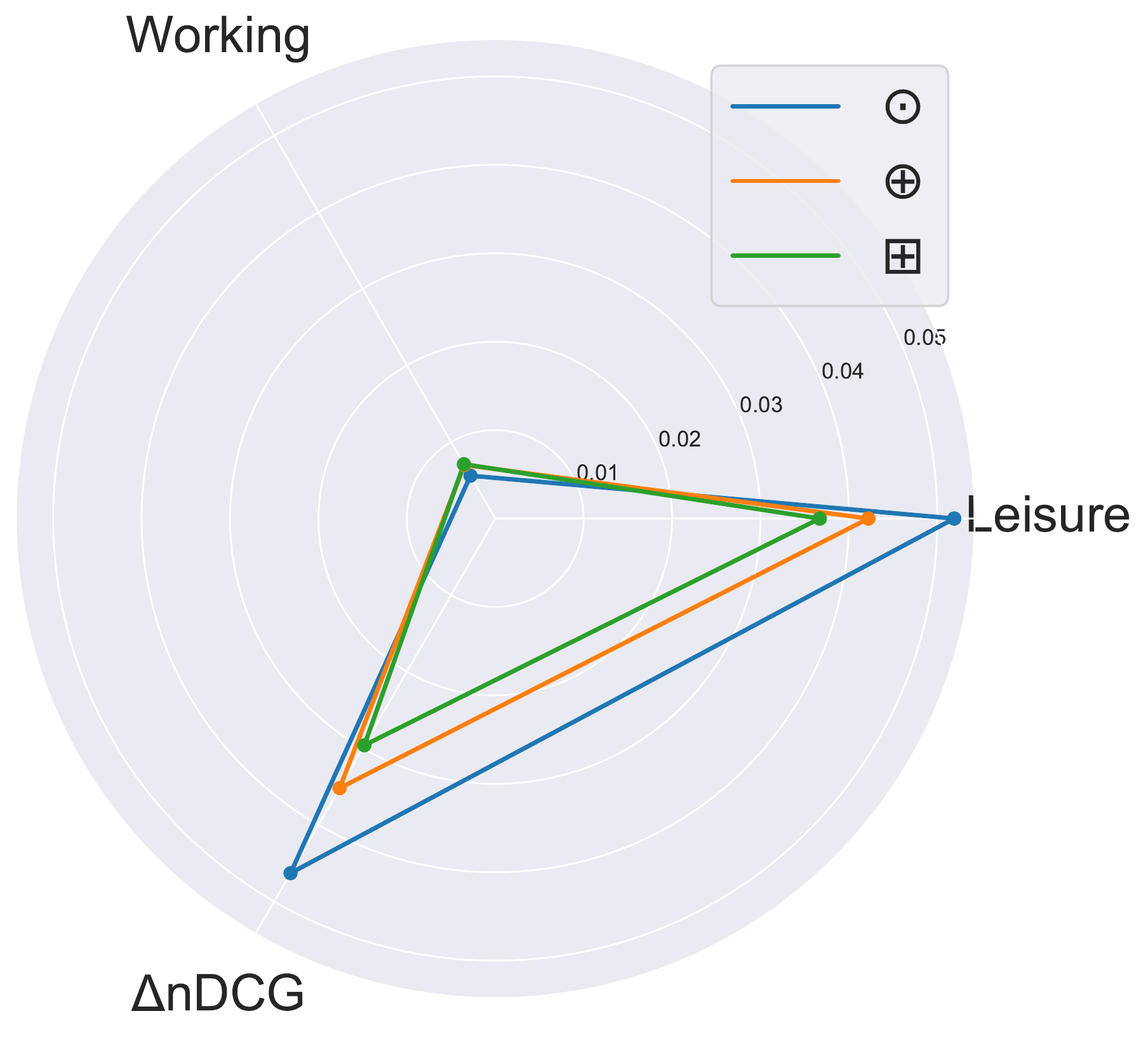}
    \label{fig:Yelp_GeoSoCa_10}
  }
    \subfloat[Yelp | LORE]
  {
    \includegraphics[scale=0.2]{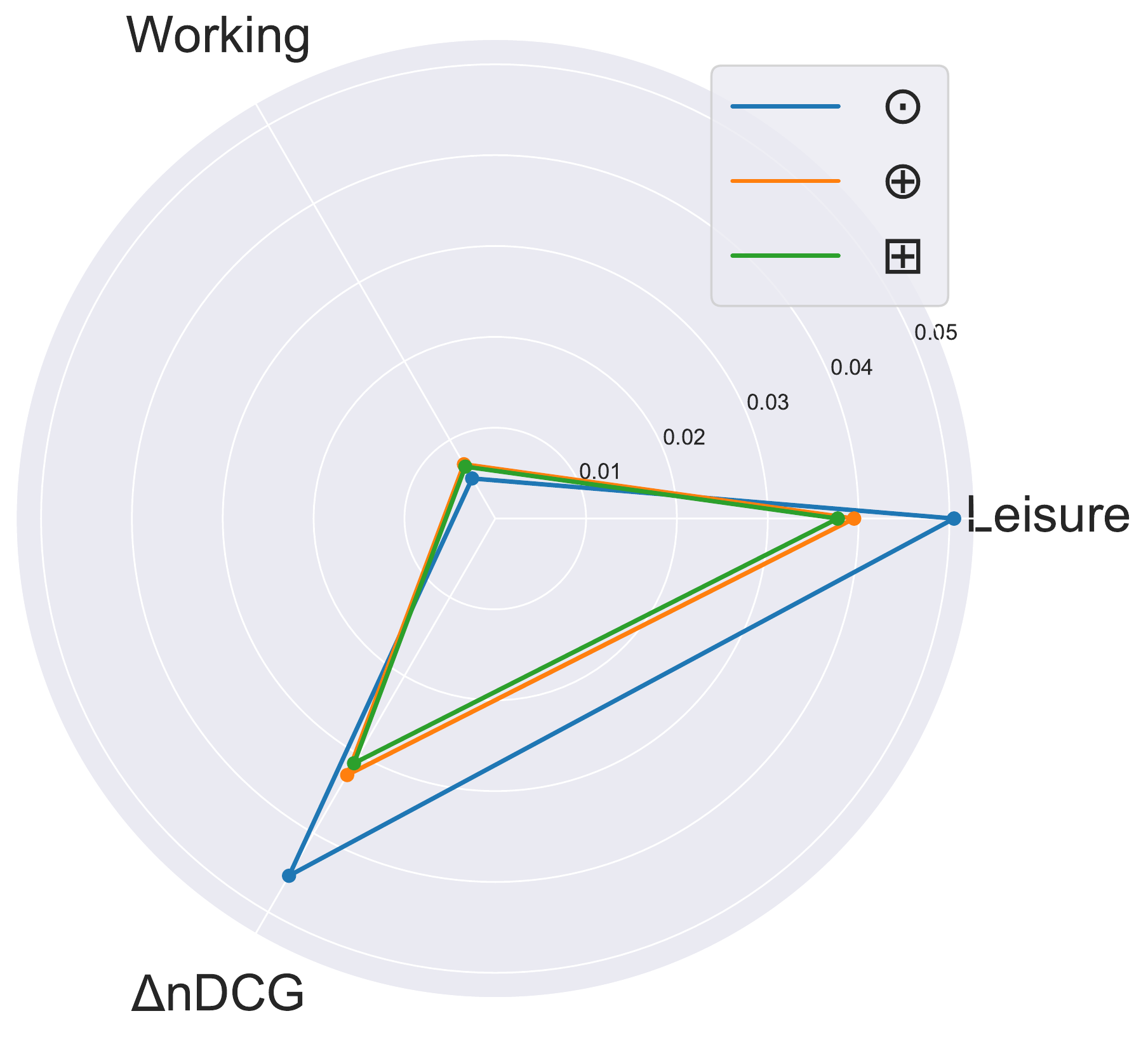}
    \label{fig:Yelp_LORE_10}
  }
  \caption{Context fusion effect on nDCG of leisure and working user groups and the unfairness between them measured by $\Delta$nDCG.}
  \label{fig:radarchart}
\end{figure*}

Fig.~\ref{fig:radarchart} illustrates that the context fusion methods would result in varying behavior in terms of accuracy for two user groups, namely, leisure and working groups. In particular, our results indicate that sum fusion ($\bigoplus$) can considerably improve the fairness criteria by sacrificing some of the model's overall accuracy in the process as the smaller area in the triangle suggest. To evaluate the cost of this improvement, the ratio of overall accuracy (nDCG) over unfairness ($\Delta$nDCG) is calculated. Having a high value in this ratio would be indicative of better overall performance, taking into account user personalization and fairness simultaneously. The average value of Acc./Unf. for ($\bigoplus$, $\bigodot$) are ($0.9145$, $0.7691$) and ($0.5049$, $0.4275$) in Gowalla and Yelp dataset, respectively. 

Moreover, we studied the effect of weighted sum fusion, $\bigboxplus$, by varying the importance of the estimated context as indicated in Eq.~\ref{eq:ployregcontext} to investigate the interplay between accuracy and fairness criteria with respect to each context. Here we limited our study to global weights of ($\lambda_1$, $\lambda_2$, $\lambda_3$), that is, the same context weight was used for all users and left the local user-level weights effect for future work. The results are plotted in Fig.~\ref{fig:linechart} in which the x-axis shows the weights and the y-axis shows the degree of fairness measured by $\Delta$nDCG (the lower the better). 
According to Fig.~\ref{fig:linechart}, the assigned weights are likely to affect the overall fairness of the context-aware models to a substantial extent. The degree of this effect, however, varies across datasets. A wider range of variation can be observed in Yelp compared to Gowalla, which may be due to the underlying data characteristics of these models, motivating the need to analyze the effects of data properties on model performance in future research~\cite{rahmani2022unfairness}.

\section{Conclusion and Future Work}
\label{sec:conclusion}
In this work, we examined the temporal unfairness of context-aware algorithms from the user's perspective. First, the analysis of the POI datasets show a temporal bias in the data, that is, the check-in distribution of users over time is not uniform, and different users have distinct preferences for checking in during leisure and work hours. Furthermore, we divided users based on their interest in check-in time into two categories of leisure-focused users and working-focused users and analyzed the characteristics of each group. In our investigation, these users have similar activity levels and consumption habits, but rather are divided by check-in behavior. Second, we show that context-aware recommender algorithms are unfair at capturing users' temporal preferences and deliver biased recommendations to both groups.

Future works considers exploring the explored temporal (and also spatial) bias in more recent approaches to POI recommendation, such as session-based models\cite{jannach2022session,knees2019recsys} and/or exploring content (e.g., images of POIs)~\cite{deldjoo2022multimedia,deldjoo2018content}. Moreover, we plan to investigate approaches that can mitigate these unfair effects, through the lens of data characteristics \cite{deldjoo2021explaining} or via algorithmic interventions~\cite{naghiaei2022cpfair}. The survey by \citet{deldjoo2022survey} provides a frame of reference for recent approaches to recommender system fairness and its multi-aspect evaluation.

\bibliographystyle{ACM-Reference-Format}
\bibliography{reference}

\end{document}